\documentclass[conference]{IEEEtran}
\ifCLASSINFOpdf
\else
\fi

\usepackage{cite}
\usepackage{amsmath,amssymb,amsfonts}
\usepackage{algorithmic}
\usepackage{graphicx}
\usepackage{epstopdf}
\usepackage{textcomp}
\usepackage{subfigure}
\usepackage{booktabs}
\usepackage{slashbox}

\def\BibTeX{{\rm B\kern-.05em{\sc i\kern-.025em b}\kern-.08em T\kern-.1667em\lower.7ex\hbox{E}\kern-.125emX}}

\allowdisplaybreaks[4]
\begin{document}

\title{Mutual Heterogeneous Signcryption Schemes for 5G Network Slicings}

\author{\IEEEauthorblockN{Jingwei Liu\IEEEauthorrefmark{1},
Lihuan Zhang\IEEEauthorrefmark{1},
Rong Sun\IEEEauthorrefmark{1},
Xiaojiang Du\IEEEauthorrefmark{2}, and
Mohsen Guizani\IEEEauthorrefmark{3}
}
\IEEEauthorblockA{\IEEEauthorrefmark{1}State Key Lab of ISN, Xidian University, Xi'an, 710071, China.\\ Email: jwliu@mail.xidian.edu.cn, zhanglihuan678@163.com, rsun@mail.xidian.edu.cn}
\IEEEauthorblockA{\IEEEauthorrefmark{2}Department of Computer and Information Sciences, Temple University, Philadelphia, PA 19122, USA.\\ Email: dxj@ieee.org}
\IEEEauthorblockA{\IEEEauthorrefmark{3}Department of Electrical and Computer Engineering, University of ldaho, Mosocow, ldaho, USA.\\ Email: mguizani@ieee.org}
}
\maketitle

\begin{abstract}
With the emerging of mobile communication technologies, we are entering the fifth generation mobile communication system (5G) era. Various application scenarios will arise in the 5G era to meet the different service requirements. Different 5G network slicings may deploy different public key cryptosystems. The security issues among the heterogeneous systems should be considered. In order to ensure the secure communications between 5G network slicings, in different public cryptosystems, we propose two heterogeneous signcryption schemes which can achieve mutual communications between the Public Key Infrastructure (PKI) and the CertificateLess public key Cryptography (CLC) environment. We prove that our schemes have the INDistinguishability against Adaptive Chosen Ciphertext Attack (IND-CCA2) under the Computational Diffie-Hellman Problem (CDHP) and the Existential UnForgeability against adaptive Chosen Message Attack (EUF-CMA) under the Discrete Logarithm Problem (DLP) in the random oracle model. We also set up two heterogeneous cryptosystems on Raspberry Pi to simulate the interprocess communication between different public key environments. Furthermore, we quantify and analyze the performance of each scheme. Compared with the existing schemes, our schemes have greater efficiency and security.
\end{abstract}
\IEEEpeerreviewmaketitle

\section{Introduction}
\label{sec:introduction}
\IEEEPARstart{S}{ince} the emergence of wireless networks, the Mobile Internet has experienced explosive growth four times (1G, 2G, 3G, 4G). It has become the foundation of information networks connecting the human society. The existing traditional communication services are difficult to adapt to many application scenarios. Therefore, the fifth generation mobile communication technology arises at this historic moment. 5G is a hot spot in global research. In 2012, the European Union officially launched METIS (mobile and wireless communications enables for the 2020 information society) project [1]. In Asia, South Korea started the ``GIGA Korea'' 5G project [2] in 2013. Chinese IMT-2020 promotion group and the ``863'' plan were also launched in June, 2013 and March, 2014 respectively [3]. Due to 5G services having the advantages of faster speed, larger capacity, and lower cost, many scholars all over the world are carrying out a wide range of research on the candidate frequency bands of 5G communication, the development of 5G vision, 5G's application requirements and key technologies [4], while, at the same time they also raise more security challenges. People who use 5G's services face more extensive and complex security threats.

\begin{figure*}[htpt]
   %Requires \usepackage{graphicx}
\centering
  \includegraphics[width=16cm]{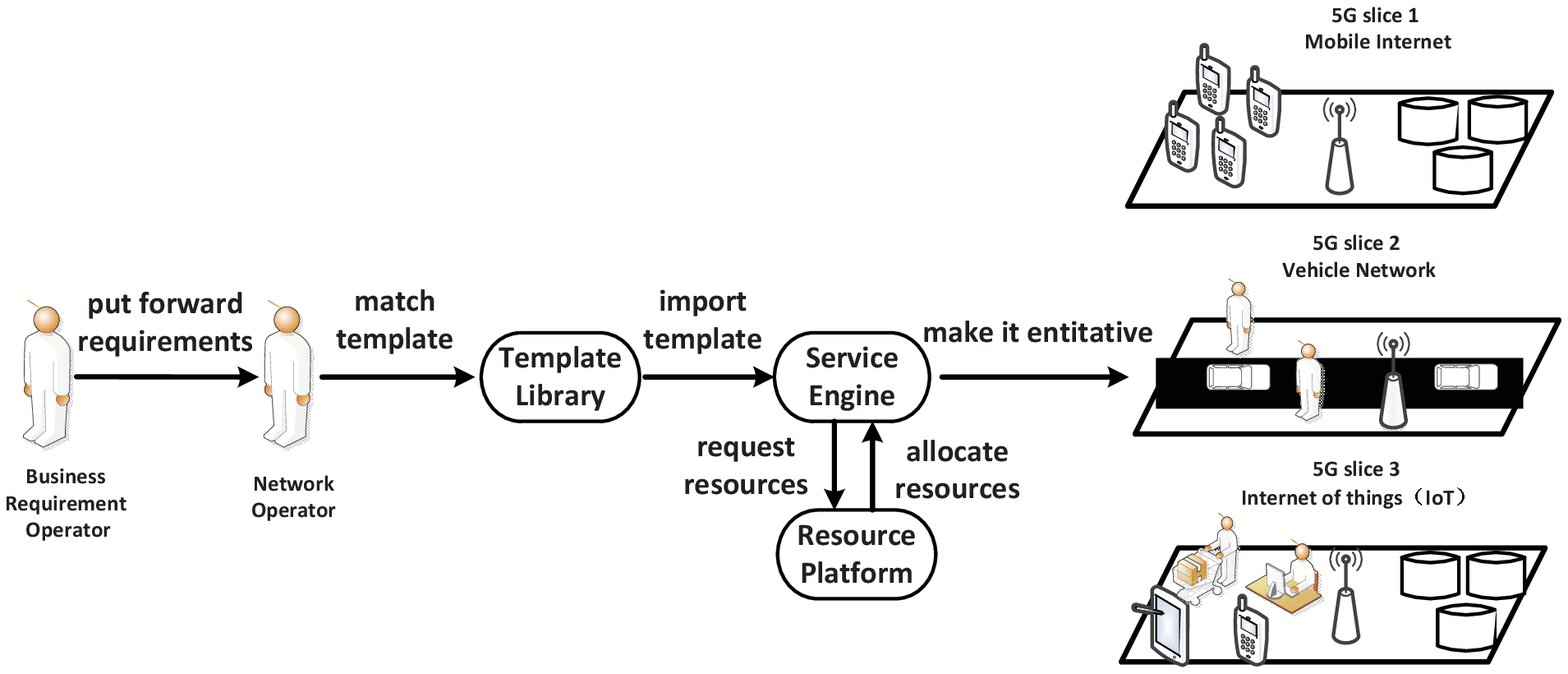}
  \caption{The system architecture of 5G}
\end{figure*}

Since a variety of 5G application scenarios may have different requirements, 5G network resources are divided into different network slicings to meet these demands. Dividing the network can reduce the network operating cost and improve the efficiency.

\subsection{Network Slicing}
The 5G system is composed of three layers shown as follows. They correlate to each other through the entities of network layout [5]:
\begin{itemize}
     \item Infrastructure Resource Layer: Through virtualization principles, the physical resources of the fixed and mobile convergence network are exposed to the orchestration entity. These resources are composed of access nodes, cloud nodes, 5G devices, network nodes and related links.
     \item Business Enablement Layer: All the functions of the convergence network should be constructed in a modular form and documented to the database. The functions of the software module and the configuration parameters of the specific network parts can be downloaded from the resource database.
     \item Business Application Layer: This layer deploys the specific applications and services in 5G network [6].
   \end{itemize}

In 2015, Ericsson proposed that 5G systems would be built to logical network slicings, which can enable operators to meet the wide range of users demands. The network slicings are also called the ``5G network slicings''. They comprise a group of network functions, resources, connection relationships and typically covering multiple technical domains including terminals, access networks, \emph{etc}. Through the virtual independent logic network infrastructure, the 5G network slicings technology provide an isolated network environment for different applications. In this way, a wide variety of scenarios can be customized according to the demands of network functions and characteristics. Different network slicings include different proprietary networks with separate logics. A 5G slicing is composed of various functions and specific Radio Access Technology (RAT) sets [7]. It can span all domains of the network: software modules running on the cloud, specific configurations of the transport network supporting flexible and accurate positioning, dedicated radio configurations, ect. The specific application scenarios or business models can decide the utilization forms of network functions and the compound modes of RAT sets.

Via the virtualization technology, network infrastructure resources are virtualized into a number of proprietary networks according to the requirements of specific applications. Slicings can customize network functions and manage network resources based on different business scenarios [8]. Each network slicing can be abstracted as a logical network formed by the collection of network functions and their corresponding configurations [9]. These logical networks (different 5G slicings) can provide network services accordingly.

There are three steps in a slicing's life cycle: the creation, the management, and the revocation. As shown in Fig. 1, the business requirement operator puts forward
requirements to the network operator, upon receiving these requirements, the network operator matches the network slicing template according to the requirements of the business scenario. A slicing template contains the network function components required,
the component interaction interfaces, and the description of the network resources. When the template is imported, the service engine can apply for network resources from the resource platform. After acquiring the resources, the service engine can use them to achieve virtual network functions and make it entitative [10]. Fig. 1 shows that 5G network slicings could be the Mobile Internet, Internet of Vehicles, Internet of Things, \emph{etc}.

\subsection{Related Work}
In a traditional PKI cryptosystem, there is a Certification Authority (CA) that issues public key certificates for each user and binds certificates with their identities. However, as the number of users increases, this method uses a great deal of time and storage space in certificate management. In order to solve this problem, Identity-Base Cryptosystem (IBC) [11] was proposed by Shamir in 1984. In IBC, users' public keys are their identities and private keys are generated by Key Generation Centers (KGC). However, it leads another issue of key escrow; because of this issue, some key management schemes have been proposed to address this problem [12, 13]. In [14], CLC was also introduced to solve this problem in which the private key is formed by two parts. One problem is the secret value of the user's choice and the other is the partial private key issued by KGC. Hence, KGC has no way to get the full private key, so the key escrow problem is solved effectively.

In order to achieve confidentiality, integrity, authentication and nonrepudiation simultaneously, a traditional approach is first to sign a message and then to encrypt it, called the signature-then-encryption approach. For the sake of optimizing algorithm and improving efficiency, the concept of signcryption [15] was first introduced by Zheng 1997, and the formal security model of signcryption [16] was first proposed in 2002. Signcryption is a new cryptographic primitive that fulfills the functions of digital signature and public key encryption in a single logical step, at a cost significantly lower than that required by the traditional signature-then-encryption approach. Although some signcryption schemes based on PKI or CLC [17, 18] were proposed in the past few years, they are only suitable for homogeneous environments. Research on heterogeneous communications has been considered by scholars around the world, in order to facilitate communications [19, 20, 21]. Similarly, in the 5G system, different network slicings may deploy different public cryptosystems. In order to ensure the secure communications between 5G network slicings over different public key systems, we put forward two innovative mutual heterogeneous signcryption schemes.

In 2010, Sun \emph{et al.} proposed a heterogeneous signcryption scheme between PKI and IBC [22] and discussed it in the multi receiver setting, however their scheme could only resist outside attacks, and it did not satisfy the non-repudiation. In 2011, Huang \emph{et al.} [23] proposed a heterogeneous signcryption scheme for PKI-to-IBC that can achieve the insider security. Regardless, it only permits a sender in the IBC to transmit a message to a receiver in the PKI and does not provide the mutual communications. In 2013, Li \emph{et al.} [24] proposed two signcryption schemes that provided bidirectional communication between PKI and IBC. This scheme relied on the cumbersome pairing operations which lead to the inefficiency. In 2016, Li \emph{et al.} [25] proposed a heterogeneous signcryption to design an access control scheme without certificates. In 2017, a concrete heterogeneous signcryption scheme of IBC-to-CLC was presented in [26]. In the same year, Wang \emph{et al.} [27] proposed an IBC-to-PKI heterogeneous signcryption scheme in the standard model.

\begin{figure*}[htpt]
   %Requires \usepackage{graphicx}
\centering
  \includegraphics[width=16cm ]{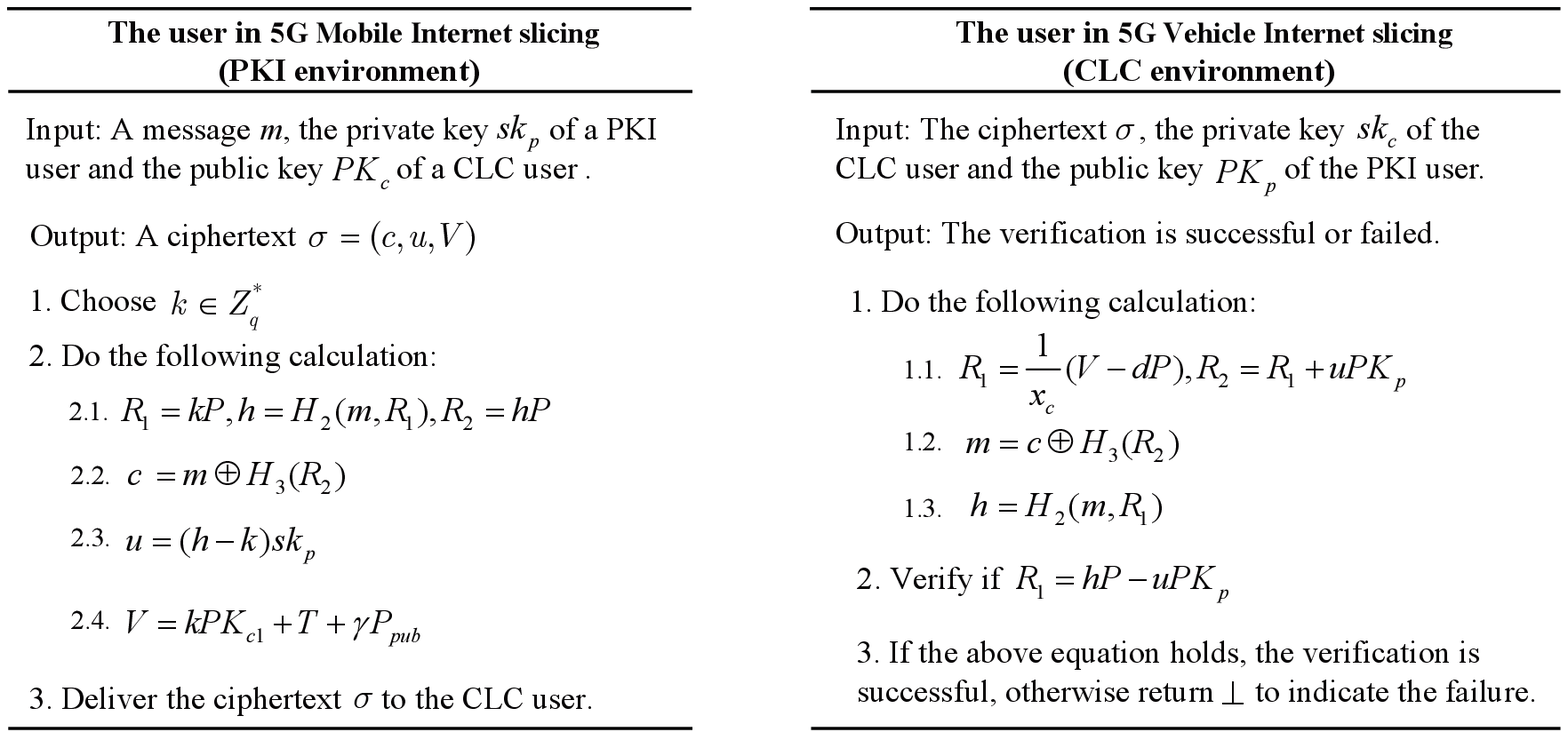}
  \caption{The PCHS scheme.}
\end{figure*}

In this paper, we propose two mutual heterogeneous signcryption schemes between PKI and CLC public cryptosystems named PCHS (PKI-CLC Heterogeneous Signcryption) and CPHS (CLC-PKI Heterogeneous Signcryption).
When the users in the 5G slicing based on PKI environment (such as a Mobile Internet slicing) try to communicate with the users in the 5G slicing based on CLC (such as a Vehicle Internet slicing), they can use the PCHS scheme to establish a secure communication, if in the opposite case, the CPHS scheme can be used.

The rest of paper is organized as follows. We introduce the preliminary work e.g. the generic signcryption model and the complexity assumptions in Section II. Then, we propose two efficient mutual heterogeneous signcryption schemes in Section III. The security analysis of the proposed schemes are given in Section IV. The performance is evaluated in Section V. Section VI concludes this paper.

\section{Preliminaries}
A heterogeneous signcryption scheme generally consists of the following five algorithms:

\emph{Setup:} This is a probabilistic algorithm running by Private Key Generator (PKG). It inputs a security parameter $l$, and outputs the system parameters and the master key. PKG publishes the system parameters while keeping the master key $s$ in secret.

\emph{PKI-KG:} This is a key generation algorithm for PKI users. Each user chooses his/her private key $SK_p$ and publishes the public key $PK_p$.

\emph{CLC-KG:} This is a key generation algorithm for CLC users.
\begin{itemize}
     \item Partial Private Key Extract: The user inputs the system parameters, the master key $s$ and his/her identity $ID$, PKG outputs the partial private key $d$ and transmits it to the user in a secure way.

     \item Set Secret Value: The user inputs an identity $ID$, and outputs a secret value $x_c$.

     \item Private Key Extract: The user inputs a partial private key $d$ and a secret value $x_c$, and outputs a full private key $SK_c$.

     \item Public Key Extract: The user inputs an identity $ID$ and the secret value $x_c$, and outputs a public key $PK_c$.
   \end{itemize}

\emph{Signcrypt:} This is a probabilistic signcryption algorithm running by a sender. It takes a message $m$, the sender's private key $SK_p$ and a receiver's public key $PK_c$, then outputs the ciphertext $\sigma$.

\emph{Unsigncrypt:} This is a deterministic unsigncryption algorithm running by a receiver. It takes the ciphertext $\sigma$, the sender's public key $PK_p$ and the receiver's secret key $SK_c$. Then, it outputs the plaintext $m$, or the symbol $\bot$ if $\sigma$ is an invalid ciphertext between the specific sender and receiver.

\subsection{Bilinear Pairings}
The bilinear pairing namely Weil pairing or Tate pairing of algebraic curves is defined as a map $e:{G_1} \times {G_1} \to {G_2}$. Here, ${G_1}$ is a cyclic additive group of a large prime order $q$. $P$ is a generator of ${G_1}$. ${G_2}$ is a cyclic multiplicative group of the same order $q$. Let $a$ and $b$ be elements in $Z_q^*$. A bilinear pairing has the following properties:
  \begin{itemize}
     \item Bilinearity: Let $e(aP,bQ) = e{(P,Q)^{ab}}$, $\forall P,Q \in {G_1}$ and $a,b \in Z_q^*$. This can be related as $\forall P,Q,S \in {G_1}$, $e(P + Q,S) = e(P,S)e(Q,S)$ and $e(P,Q + S) = e(P,Q)e(P,S)$.

     \item Nondegeneracy: There exist $P,Q \in {G_1}$, such that $e\left( {P,Q} \right) \ne {1_{{G_2}}}$. Here, ${1_{{G_2}}}$ denotes the identity element of ${G_2}$.

     \item Computability: There is an efficient algorithm to compute $e\left( {P,Q} \right)$ for all $P,Q \in {G_1}$.
   \end{itemize}

\begin{figure*}[htpt]
   %Requires \usepackage{graphicx}
\centering
  \includegraphics[ width=16cm ]{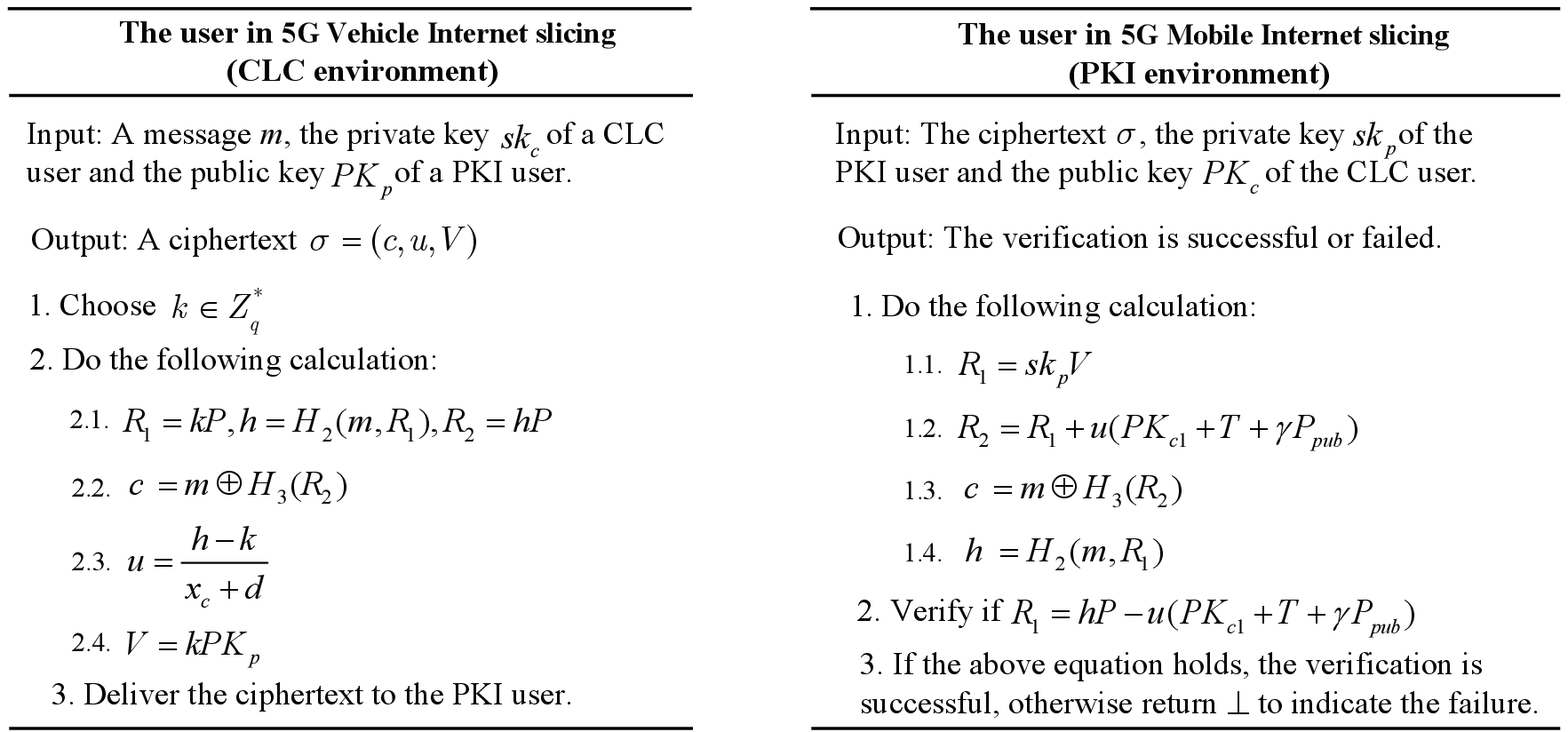}
  \caption{The CPHS scheme.}
\end{figure*}
\subsection{Complexity Assumptions}
The security of our schemes relies on the hardness of the following problems.

${G_1}$ is a cyclic additive group of a large prime order $q$. $P$ is a generator of ${G_1}$.

\emph{Definition 1.} Computational Diffie-Hellman Problem (CDHP): Given an instance $(P,aP,bP) \in {G_1}$, for any $a,b\in Z_q^*$, it is difficult to compute $abP \in {G_1}$.

\emph{Definition 2.} Discrete Logarithm Problem (DLP): Given an instance $(P,aP) \in {G_1}$, it is difficult to compute the integer $a \in Z_q^*$.

We assume there are two types of adversaries with different capabilities. The Type I adversary can replace users' public keys, but it does not know the master secret key $s$. The Type II adversary can access the master secret key $s$, but can not replace users' public keys.
\section{Proposed Schemes}
In this section, we assume that a 5G Mobile Internet slicing is in PKI public cryptosystem, and a 5G Vehicle Internet slicing is in CLC public cryptosystem. We propose two efficient signcryption schemes for the security authentication between the two heterogeneous 5G slicings. The first scheme PCHS allows users in PKI cryptosystem to signcrypt the messages and send them to users in CLC cryptosystem. Upon receiving these signcrypted messages, the users in CLC cryptosystem can decrypt and verify them. The second scheme CPHS is the inverse of PCHS. The detailed processes are shown in Fig. 2 and Fig. 3.
\subsection{PCHS}
The PCHS is described as follows:

\emph{Setup:} Given a security parameter $l$, PKG chooses a cyclic additive group ${G_1}$ of a large prime order $q$. $P$ is a generator of ${G_1}$. Then, PKG defines three cryptographic hash functions: ${H_1}:{\{ 0,1\} ^*} \times {G_1} \to Z_q^*$, ${H_2}:{\{ 0,1\} ^n} \times {G_1} \to Z_q^*$, and ${H_3}:{G_1} \to {\{ 0,1\} ^n}$. PKG selects a master secret key $s \in Z_q^*$ randomly and computes the master public key ${P_{pub}} = sP$. Then, it publishes system parameters $\{ {G_1},P,{P_{pub}},n,l,{H_1},{H_2},{H_3}\}$ and keeps the master key $s$ secret.

\emph{PKI-KG:} A user in PKI cryptosystem chooses a random number ${x_p} \in Z_q^*$ as his/her private key $s{k_p}$ and computes $\frac{1}{{{x_p}}}P$ as his/her public key $P{K_p}$.

\emph{CLC-KG:}
\begin{itemize}
     \item Partial Private Key Extract: PKG randomly selects $t \in Z_q^*$ and computes $T = tP$, $\gamma = {H_1}(ID,T)$, and $d = t + s\gamma \in Z_q^*$. Then, PKG sends $d$ and $T$ to the user securely. $d$ is the user's partial private key.

     \item Set Secret Value: The user randomly chooses ${x_c} \in Z_q^*$ as his/her secret value.

     \item Private Key Extract: The user sets his/her full private key as $s{k_c} = ({x_c},d)$.

     \item Public Key Extract: The user sets his/her public key as $P{K_c} = (T,P{K_{c1}} = {x_c}P)$.
   \end{itemize}

\emph{Signcrypt:} The user in PKI cryptosystem uses his/her private key $s{k_p}$ and the receiver's public key $P{K_c}$ in CLC cryptosystem to signcrypt a message $m$ as follows:

1) Choose a number $k \in Z_q^*$ randomly.

2) Compute ${R_1} = {kP}$, $h = {H_2}(m,{R_1})$, ${R_2} = {hP}$.

3) Compute $c = m \oplus {H_3}({R_2})$.

4) Compute $u = (h - k)s{k_p}$.

5) Compute $V = kP{K_{c1}} + T + \gamma {P_{pub}}$.

The ciphertext is $\sigma  = (c,u,V)$.

\emph{Unsigncrypt:} The user in CLC cryptosystem uses his/her private key $s{k_c}$ and the sender's public key $P{K_p}$ in PKI cryptosystem to unsigncrypt the ciphertext $\sigma  = (c,u,V)$ as follows:

1) Compute ${R_1} = \frac{1}{{{x_c}}}(V - dP)$.

2) Compute ${R_2} = {R_1} + uP{K_p}$.

3) Compute $m = c \oplus {H_3}({R_2})$.

4) Compute $h \;= {H_2}(m,{R_1})$.

5) Accept the message if and only if ${R_1} = hP - uP{K_p}$,
return $\bot$ otherwise.

Now we verify the correctness of the PCHS.

Firstly:
\[\begin{array}{l}
{\kern 1pt} {\kern 1pt} {\kern 1pt} {\kern 1pt} {\kern 1pt} {\kern 1pt} {\kern 1pt} {\kern 1pt} {\kern 1pt} {\kern 1pt} \frac{1}{{{x_c}}}(V - dP)\\
 = \frac{1}{{{x_c}}}(kP{K_{c1}} + T + \gamma {P_{pub}} - dP)\\
 = \frac{1}{{{x_c}}}(k{x_c}P + tP + \gamma sP - dP)\\
 = \frac{1}{{{x_c}}}(k{x_c}P + dP - dP)\\
 = \frac{1}{{{x_c}}} \cdot k{x_c}P\\
 = kP = {R_1}
\end{array}\]
Secondly:
\[\begin{array}{l}
{\kern 1pt} {\kern 1pt} {\kern 1pt} {\kern 1pt} {\kern 1pt} {\kern 1pt} {\kern 1pt} {\kern 1pt} {\kern 1pt} {R_1} + uP{K_p}\\
 = {R_1} + (h - k)s{k_p} \cdot \frac{1}{{{x_p}}}P\\
 = kP + (h - k){x_p} \cdot \frac{1}{{{x_p}}}P\\
 = kP + hP - kP\\
 = hP = {R_2}
\end{array}\]
Finally:
\[\begin{array}{l}
{\kern 1pt} {\kern 1pt} {\kern 1pt} {\kern 1pt} {\kern 1pt} {\kern 1pt} {\kern 1pt} {\kern 1pt} {\kern 1pt} hP - uP{K_p}\\
 = hP - (h - k){x_p} \cdot \frac{1}{{{x_p}}}P\\
 = hP - hP + kP\\
 = kP = {R_1}
\end{array}\]
\subsection{CPHS}
In CPHS, the \emph{Setup}, \emph{PKI-KG} and \emph{CLC-KG} algorithms are the same as PCHS. And CPHS can be described as follows:

\emph{Signcrypt:} The user in CLC cryptosystem uses his/her private key $s{k_c}$ and the receiver's public key $P{K_p}$ in PKI cryptosystem to signcrypt a message $m$ as follows:

1) Choose a number $k \in Z_q^*$ randomly.

2) Compute ${R_1} = {kP}$, $h = {H_2}(m,{R_1})$, ${R_2} = {hP}$.

3) Compute $c = m \oplus {H_3}({R_2})$.

4) Compute $u = \frac{{h - k}}{{{x_c} + d}}$.

5) Compute $V = kP{K_p}$.

The ciphertext is $\sigma  = (c,u,V)$.

\emph{Unsigncrypt:} The user in PKI cryptosystem uses his/her private key $s{k_p}$ and the sender's public key $P{K_c}$ in CLC cryptosystem to unsigncrypt the ciphertext $\sigma  = (c,u,V)$ as follows:

1) Compute ${R_1} = s{k_p}V$.

2) Compute ${R_2} = {R_1} + u(P{K_{c1}} + T + \gamma {P_{pub}})$.

3) Compute $m = c \oplus {H_3}({R_2})$.

4) Compute $h\; = {H_2}(m,{R_1})$.

5) Accept the message if and only if ${R_1} = hP - u(P{K_{c1}} + T + \gamma {P_{pub}})$,
return $\bot$ otherwise.

Now we verify the correctness of the CPHS.

Firstly:
\[\begin{array}{l}
s{k_p}V {\kern 1pt} {\kern 1pt} = ks{k_p}P{K_p}\\
{\kern 1pt} {\kern 1pt} {\kern 1pt} {\kern 1pt} {\kern 1pt} {\kern 1pt} {\kern 1pt} {\kern 1pt} {\kern 1pt} {\kern 1pt} {\kern 1pt} {\kern 1pt} {\kern 1pt} {\kern 1pt} {\kern 1pt} {\kern 1pt} {\kern 1pt} {\kern 1pt} {\kern 1pt} {\kern 1pt} {\kern 1pt} {\kern 1pt} {\kern 1pt} {\kern 1pt} {\kern 1pt}  = k\frac{1}{{{x_p}}} \cdot {x_p}P\\
{\kern 1pt} {\kern 1pt} {\kern 1pt} {\kern 1pt} {\kern 1pt} {\kern 1pt} {\kern 1pt} {\kern 1pt} {\kern 1pt} {\kern 1pt} {\kern 1pt} {\kern 1pt} {\kern 1pt} {\kern 1pt} {\kern 1pt} {\kern 1pt} {\kern 1pt} {\kern 1pt} {\kern 1pt} {\kern 1pt} {\kern 1pt} {\kern 1pt} {\kern 1pt} {\kern 1pt} {\kern 1pt}  = kP = {R_1}
\end{array}\]
Secondly:
\[\begin{array}{l}
{\kern 1pt} {\kern 1pt} {\kern 1pt} {\kern 1pt} {\kern 1pt} {\kern 1pt} {\kern 1pt} {\kern 1pt} {\kern 1pt} {\kern 1pt} {\kern 1pt} {R_1} + u(P{K_{c1}} + T + \gamma {P_{pub}}){\kern 1pt} {\kern 1pt} \\
 = kP + \frac{{h - k}}{{{x_c} + d}} \cdot (P{K_{c1}} + T + \gamma {P_{pub}})\\
 = kP + \frac{{h - k}}{{{x_c} + d}} \cdot ({x_c}P + tP + \gamma sP)\\
 = kP + \frac{{h - k}}{{{x_c} + d}} \cdot ({x_c} + d)P\\
 = kP + (h - k)P\\
{\rm{ = }}{\kern 1pt} {\kern 1pt} hP{\kern 1pt} {\kern 1pt} {\rm{ =  }}{R_{\rm{2}}}
\end{array}\]

Finally:
\[\begin{array}{l}
{\kern 1pt} {\kern 1pt} {\kern 1pt} {\kern 1pt} {\kern 1pt} {\kern 1pt} {\kern 1pt} {\kern 1pt} {\kern 1pt} hP - u(P{K_{c1}} + T + \gamma {P_{pub}}){\kern 1pt} {\kern 1pt} \\
 = hP - \frac{{h - k}}{{{x_c} + d}} \cdot (P{K_{c1}} + T + \gamma {P_{pub}})\\
 = hP - \frac{{h - k}}{{{x_c} + d}} \cdot ({x_c}P + tP + \gamma sP)\\
 = hP - \frac{{h - k}}{{{x_c} + d}} \cdot ({x_c} + d)P\\
 = hP - (h - k)P\\
{\rm{ = }}{\kern 1pt} {\kern 1pt} kP{\kern 1pt} {\kern 1pt} {\rm{ =  }}{R_1}
\end{array}\]

\section{The Security Proof of Schemes}
In this section, we prove the security of the proposed schemes.

\noindent{\textbf{Theorem 1. (PKI-CLC IND-CCA2-1):}} In the random oracle model [28], if an adversary ${{\cal A}_I}$ has a nonnegligible advantage $\xi $ against the IND-CCA2-1 security of the PCHS when performing ${q_i}$ queries to oracles ${H_i}$ $(i = 1,2,3)$, there exists an algorithm through which the challenger ${\cal C}$ can solve the CDHP with an advantage ${\xi ^*}$. Here, ${\xi ^*} \ge \frac{\xi }{{{q_1}^2}}(1 - \frac{{{q_{e1c}}}}{{{q_1}}})(1 - \frac{{{q_{e2c}}}}{{{q_1}}})(1 - \frac{{{q_u}}}{{{2^l}}})$, ${q_{{e_{1c}}}}$ denotes partial private key queries of CLC system, ${q_{{e_{2c}}}}$ denotes private key queries of CLC system, and ${q_{u}}$ denotes unsigncryption queries.

\noindent{\textbf{Proof:}} To solve a random CDHP instance $(P,aP,bP)$, ${\cal C}$ uses ${{\cal A}_I}$ as a subroutine. ${{\cal A}_I}$ should ask for ${H_1}$ before the identity $I{D_i}$ is used in any other queries.

\emph{\textbf{Initialization:}} ${\cal C}$ initializes and returns the system parameters $\{ {G_1},P,{P_{pub}},n,l,{H_1},{H_2},{H_3}\}$ to ${{\cal A}_I}$. Next, it picks a challenged identity $\alpha  \in \{ 1,2, ..., {q_1}\} $ randomly without leaking anything about $I{D_\alpha }$ to ${{\cal A}_I}$. ${\cal C}$ needs to maintain the list of ${L_1}$\textasciitilde${L_3}$, $L{K_p}$ and $L{K_c}$ that are used to simulate the ${H_1}$\textasciitilde${H_3}$ hash oracles and the key extraction oracles respectively. Then, it answers these queries as follows.

\emph{\textbf{Phase 1:}} ${{\cal A}_I}$ performs a polynomially bounded number of queries in an adaptive manner.

\quad 1. \emph{${H_1}$-Queries:} When ${{\cal A}_I}$ presents this query on an identity $I{D_i}$, ${\cal C}$ checks whether the tuple $(I{D_i},{T_i},{\gamma _i})$ exists in ${L_1}$. If so, ${\cal C}$ returns ${\gamma _i}$ to ${{\cal A}_I}$. Otherwise, the public key query on $I{D_i}$ is made to generate ${\gamma _i}$ to ${{\cal A}_I}$ subsequently.

\quad 2. \emph{${H_2}$-Queries:} For a ${H_2}$ query, ${\cal C}$ first checks if $({m_i},{R_{1i}},{h_{i}})$ has been in ${L_2}$ previously. If so, ${\cal C}$ returns the value $h_{i}$. Otherwise, ${\cal C}$ returns a random $h_{i} \in Z_q^*$, and adds new tuple $({m_i},{R_{1i}},{h_{i}})$ to ${L_2}$.

\quad 3. \emph{${H_3}$-Queries:} For a ${H_3}$ query, ${\cal C}$ first checks if the value of ${H_3}$ query has been in ${L_3}$ previously. If so, ${\cal C}$ returns it. Otherwise, ${\cal C}$ randomly chooses ${h_{3i}}$ from ${\{ 0,1\} ^n}$, returns ${h_{3i}}$ and adds $({R_{2i}},{h_{3i}})$ to ${L_3}$.

\quad 4. \emph{PKI Private-Key-Queries:} For a private key query on $I{D_i}$, ${\cal C}$ will invoke $L{K_p}$ and search $(I{D_i},{x_{pi}},P{K_{pi}})$, then returns the private key ${x_{pi}}$.

\quad 5. \emph{CLC Partial-Private-Key-Queries:} For a partial private key query on $I{D_i}$, ${\cal C}$ makes the following response:

\quad (1) If $I{D_i} = I{D_\alpha }$, ${\cal C}$ aborts.

\quad (2) If $I{D_i} \ne I{D_\alpha }$, ${\cal C}$ will invoke $L{K_c}$ and search $(I{D_i},{x_{ci}},{d_i},{T_i},P{K_{c1i}})$, then return the partial private key ${d_i}$.

\quad 6. \emph{CLC Private-Key-Queries:} For a private key query above $I{D_i}$, ${\cal C}$ responds as follows:

\quad (1) If $I{D_i} = I{D_\alpha }$, ${\cal C}$ aborts.

\quad (2) If $I{D_i} \ne I{D_\alpha }$, ${\cal C}$ will invoke $L{K_c}$ and search $(I{D_i},{x_{ci}},{d_i},{T_i},P{K_{c1i}})$, then return $({x_{ci}},{d_i})$.

\quad 7. \emph{CLC Public-Key-Queries:} When ${{\cal A}_I}$ submits a query on identity $I{D_i}$ for his/her public key, ${\cal C}$ searches $L{K_c}$ for the tuple $(I{D_i},{x_{ci}},{d_i},{T_i},P{K_{c1i}})$ and responds as follows:

\quad (1) If the public key exists, ${\cal C}$ returns $({T_i},P{K_{c1i}})$.

\quad (2) If $I{D_i} = I{D_\alpha }$, ${\cal C}$ randomly chooses ${x_{c\alpha }}, {\gamma _\alpha } \in Z_q^*$, $bP \in {G_1}$ with unknown $b$, then it sets $P{K_{c1}}_\alpha = bP$, ${T_\alpha } = {d_{i\alpha }}P - {\gamma _\alpha }{P_{pub}}$, where ${\gamma _\alpha } = {H_1}(I{D_\alpha },{T_\alpha })$. ${\cal C}$ updates $(I{D_\alpha },{x_{c\alpha }},{d_{i\alpha }},{T_\alpha },P{K_{c1\alpha }})$ in $L{K_c}$ and $(I{D_\alpha },{T_\alpha },{\gamma _\alpha })$ in ${L_1}$.

\quad (3) If $I{D_i} \ne I{D_\alpha }$, ${\cal C}$ selects ${\gamma _i},{x_{ci}},{d_i} \in Z_q^*$ randomly. Next, ${\cal C}$ calculates $P{K_{c1}}_i = bP$, ${T_i} = {d_i}P - {\gamma _i}{P_{pub}}$, where ${\gamma _i} = {H_1}(I{D_i},{T_i})$. Then, ${\cal C}$ updates $L{K_c}$ with $(I{D_i},{x_{c,i}},{d_i},{T_i},P{K_{c1i}})$ and ${L_1}$ with $(I{D_i},{T_i},{\gamma _i})$.

\quad 8. \emph{CLC Public-Key-Replacement-Queries:} When ${{\cal A}_I}$ replaces the public key $P{K_{ci}}$ of the identity $I{D_i}$ with $P{K_{ci}}^*$, ${\cal C}$ updates $L{K_c}$ with the tuple $(I{D_i},\emptyset,\emptyset,{T_i},P{K_{ci}}^*)$. Here, $\emptyset$ denotes an unknown value. The new public key is used by the challenger to solve the CDHP or is requested by the adversary in the public key queries.

\quad 9. \emph{Signcrypt-Queries:} Suppose $I{D_{send}}$ and $I{D_{receive}}$ are the identities of a sender and a receiver respectively. When ${{\cal A}_I}$ makes this query on
the tuple $(\sigma ,I{D_{send}},I{D_{receive}})$, ${\cal C}$ makes responses as below:

\quad (1) If $I{D_{send}} \ne I{D_\alpha }$, ${\cal C}$ runs the signcrypt algorithm normally and sends the ciphertext to ${{\cal A}_I}$.

\quad (2) If $I{D_{send}} = I{D_\alpha }$ and $I{D_{receive}} \ne I{D_\alpha }$, ${\cal C}$ finds $({d_\alpha },{x_{c\alpha }},P{K_p})$ from $L{K_c}$ and $L{K_p}$ and generates the ciphertext in following steps:
\begin{itemize}
  \item Choose $k \in Z_q^*$ randomly, compute ${R_1} = kP$;
  \item Compute $h = {H_2}(m,{R_1})$, ${R_2} = {hP}$, store $(m,{R_1},h)$ into ${L_2}$;
  \item Compute $c = m + {H_3}{\rm{(}}{R_2}{\rm{)}}$;
  \item Compute $u = \frac{{h - k}}{{{x_c} + d}}$, $V = kP{K_p}$;
  \item The ciphertext is $\sigma  = (c,u,V)$.
\end{itemize}

\quad 10. \emph{Unsigncrypt-Queries:} Upon receiving a unsigncrypt query of $(\sigma ,I{D_{send}},I{D_{receive}})$, ${\cal C}$ responds as follows:

\quad (1) If $I{D_{receive} \ne I{D_\alpha }}$, ${\cal C}$ runs the unsigncrypt algorithm normally and returns the result.

\quad (2) If $I{D_{receive}} = I{D_\alpha }$, ${\cal C}$ searches ${L_3}$ for $h_3$. ${\cal C}$ computes $m{\rm{ = }}c \oplus {h_3}$. If and only if ${R_1} = hP - uP{K_p}$, the message is accepted. Otherwise, the ciphertext is rejected.

\emph{\textbf{Challenge phase:}} ${{\cal A}_I}$ generates two equal-length plaintexts ${m_0},{m_1}$ and two challenged identities $I{D_{send}}^*$, $I{D_{receive}}^*$. In phase 1, the public key of $I{D_{receive}}^*$ can not be replaced and the partial private key queries can not be asked as well as the secret value. If $I{D_{receive}}^* \ne I{D_\alpha }$, ${\cal C}$ aborts. Otherwise, ${\cal C}$ asks public key request oracle on $I{D_{send}}^*$, sets receiver's partial public key to ${bP}$ (an instance of CDHP) for an unknown ${b}$. Next, ${\cal C}$ randomly selects $T \in {G_1}$, ${k^*} \in Z_q^*$, ${\gamma^*}, h_3^* \in {\{ 0,1\} ^n}$ and a random bit $\mu  \in \{ 0,1\}$, then sets $R_1^* = aP$, $P{K_{c1}}^* = bP$, ${c^*} = {m_\mu } \oplus h_3^*$, computes ${V^*} = {k^*}P{K_{c1}}^* + T + {\gamma ^*}{P_{pub}}$ and returns $({c^*}, {u^*}, {V^*})$ to ${{\cal A}_I}$.

\emph{\textbf{Phase 2:}} As in Phase 1, ${{\cal A}_I}$ will also present adaptively queries with the limitations of Type I adversary in this phase. However, (1) ${{\cal A}_I}$ cannot submit the private key query on $I{D_{receive}}^*$. (2) ${{\cal A}_I}$ cannot ask the partial private key query on $I{D_{send}}^*$ if the public key is replaced before the challenge. (3) ${{\cal A}_I}$ cannot present the unsigncrypt query on $({\sigma^*},I{D_{send}}^*,I{D_{receive}}^*)$.

\emph{\textbf{Guess:}} In order to get a correct guess, ${{\cal A}_I}$ should obtain the outputs $R_1^* = aP$, $P{K_{c1}}^* = bP$ and ${V^*} = {k^*}P{K_{c1}}^* + T + {\gamma ^*}{P_{pub}}$ from the challenge phase. Given an instance $(P,aP,bP)$, ${\cal C}$ can solve CDHP: $abP = {V^*} - T - {\gamma ^*}{P_{pub}}$.

\emph{\textbf{Probability Analysis:}} In above discussions, there are four situations leading to the aborting of ${\cal C}$:
\begin{itemize}
  \item $E_1$: ${{\cal A}_I}$ asks the partial private key queries of the challenged identity $I{D_\alpha }$.
  \item $E_2$: ${{\cal A}_I}$ presents the private key queries of the challenged identity $I{D_\alpha }$.
  \item $E_3$: ${{\cal A}_I}$ does not choose $I{D_\alpha }$ as the receiver's identity in the challenge phase.
  \item $E_4$: ${\cal C}$ aborts in an unsigncryption query because of rejecting a valid ciphertext.
\end{itemize}

Only if ${\cal C}$ does not reject the game, the CDHP can be solved. As a result, the probability of ${\cal C}$ not aborting is: $\Pr [\neg abort] = \Pr [\neg {E_1} \wedge \neg {E_2} \wedge \neg {E_3} \wedge \neg {E_4}]$.

We know that $\Pr [{E_1}] = \frac{{{q_{e1c}}}}{{{q_1}}}$, $\Pr [{E_2}] = \frac{{{q_{e2c}}}}{{{q_1}}}$, $\Pr [{E_3}] = (1-\frac{1}{{{q_1}}})$, and $\Pr [{E_4}] \le \frac{{{q_u}}}{{{2^l}}}$. Therefore, $\Pr [\neg abort] \ge \frac{1}{{{q_1}}}(1 - \frac{{{q_{e1c}}}}{{{q_1}}})(1 - \frac{{{q_{e2c}}}}{{{q_1}}})(1 - \frac{{{q_u}}}{{{2^l}}})$.

In addition, the probability that ${\cal C}$ randomly chooses a $T$ from ${L_1}$ and outputs it as a solution of CDHP is $\frac{1}{{{q_1}}}$. In conclusion, we have
${\xi ^*} \ge \frac{\xi }{{{q_1}^2}}(1 - \frac{{{q_{e1c}}}}{{{q_1}}})(1 - \frac{{{q_{e2c}}}}{{{q_1}}})(1 - \frac{{{q_u}}}{{{2^l}}})$.

\noindent{\textbf{Theorem 2. (PKI-CLC IND-CCA2-2):}} In the random oracle model, if an adversary ${{\cal A}_{II}}$ has a nonnegligible advantage $\xi $ against the IND-CCA2-2 security of the PCHS when performing ${q_i}$ queries to oracles ${H_i}$ $(i = 1,2,3)$, then there exists an algorithm through which the challenger ${\cal C}$ can solve the CDHP with an advantage ${\xi ^*}$. Here, ${\xi ^*} \ge \frac{\xi }{{{q_1}^2}}(1 - \frac{{{q_{e2c}}}}{{{2^l}}})(1 - \frac{{{q_u}}}{{{2^l}}})$.

The ${H_1}$\textasciitilde${H_3}$ hash oracles and the key extraction oracles are similar with PKI-CLC IND-CCA2-1. The second type adversary ${{\cal A}_{II}}$ knows the master key of PKG, but it is not allowed to replace any user's public key.

\emph{\textbf{Probability Analysis:}} In Phase 2, ${{\cal A}_{II}}$ can not ask a private key query on $ID_{receive}^*$ and an unsigncryption query on ${\sigma ^*}$ for the corresponding plaintext. ${\cal C}$ will abort the game in following situations:
\begin{itemize}
  \item $E_1$: ${{\cal A}_{II}}$ asks the private key queries of the challenged identity $I{D_\alpha }$ .
  \item $E_2$: ${{\cal A}_{II}}$ does not choose $I{D_\alpha }$ as the receiver's identity in the challenge phase.
  \item $E_3$: ${\cal C}$ aborts in an unsigncryption query because of rejecting a valid ciphertext.
\end{itemize}

Only if ${\cal C}$ does not reject the game, the CDHP can be solved. As a result, the probability of ${\cal C}$ not aborting is: $\Pr [\neg abort] = \Pr [\neg {E_1} \wedge \neg {E_2} \wedge \neg {E_3}]$.

We know that $\Pr [{E_1}] = \frac{{{q_{e2c}}}}{{{q_1}}}$, $\Pr [{E_2}] = (1-\frac{1}{{{q_1}}})$, and $\Pr [{E_3}] \le \frac{{{q_u}}}{{{2^l}}}$. Therefore, $\Pr [\neg abort] \ge \frac{1}{{{q_1}}}(1 - \frac{{{q_{e2c}}}}{{{q_1}}})(1 - \frac{{{q_u}}}{{{2^l}}})$.

In addition, the probability that ${\cal C}$ randomly chooses a $T$ from ${L_1}$ and outputs it as a solution of CDHP is $\frac{1}{{{q_1}}}$. In conclusion, we have
${\xi ^*} \ge \frac{\xi }{{{q_1}^2}}(1 - \frac{{{q_{e2c}}}}{{{q_1}}})(1 - \frac{{{q_u}}}{{{2^l}}})$.

\noindent{\textbf{Theorem 3. (PKI-CLC EUF-CMA):}} In the random oracle model, if an adversary ${\cal F}$ has a nonnegligible advantage $\xi $ against the EUF-CMA security of the PCHS when performing ${q_i}$ queries to oracles ${H_i}$ $(i = 1,2,3)$, then there exists an algorithm through which the challenger ${\cal C}$ can solve the DLP with an advantage ${\xi ^*}$. Here, ${\xi ^*} \ge \frac{\xi }{{{q_1}^2}}(1 - \frac{{{q_{ep}}}}{{{2^l}}})(1 - \frac{{{q_s}}}{{{2^l}}})$, ${q_{{ep}}}$ denotes private key queries of PKI system and ${q_s}$ denotes signcryption queries.

\noindent{\textbf{Proof:}} To solve a random DLP instance $(P,aP)$, ${\cal C}$ uses ${\cal F}$ as a subroutine. ${\cal F}$ should ask for ${H_1}$ before the identity $I{D_i}$ is used in any other queries.

\emph{\textbf{Initialization:}} ${\cal C}$ initializes and returns the system parameters $\{ {G_1},P,{P_{pub}},n,l,{H_1},{H_2},{H_3}\}$ to ${\cal F}$. Next, it picks a challenged identity $\alpha  \in \{ 1,2, ..., {q_1}\} $ randomly without leaking anything about $I{D_\alpha }$ to ${\cal F}$. ${\cal C}$ needs to maintain the list of ${L_1}$\textasciitilde${L_3}$, $L{K_p}$ and $L{K_c}$ that are used to simulate the ${H_1}$\textasciitilde${H_3}$ hash oracles and the key extraction oracles respectively.

\emph{\textbf{Training:}} ${\cal F}$ performs a polynomially bounded number of queries in an adaptive manner. The queries in this phase are the same as the queries described in Theorem 1.

\emph{\textbf{Forgery:}} After the training, ${\cal F}$ outputs a forgery $({\sigma}^*,I{D_{send}}^*,I{D_{receive}}^*)$. During the training, ${\cal F}$ cannot make a private key query on $I{D_{send}}^*$. If $I{D_{send}}^* \ne I{D_\alpha }$, ${\cal C}$ aborts. If $I{D_{send}}^* = I{D_\alpha }$, ${\cal C}$ invokes the ${L_1}$ and $L{K_c}$ to search ${\gamma*}$, ${T^*}$, ${P{K_{c1}}^*}$ and $R_1$. Then, ${\cal C}$ obtains $aP{K_{c1}} = V - T - \gamma {P_{pub}}$ and $aP{K_{c1}}^* = V - {T^*} - {\gamma ^*}{P_{pub}}$. Finally, ${\cal C}$ outputs $a = \frac{{[(t - {t^*}) + s(\gamma  - {\gamma ^*})]}}{{({x_c} - {x_c}^*)}}$ as the solution of DLP. The proof is as follows:
\[\begin{array}{l}
{\kern 1pt} {\kern 1pt} {\kern 1pt} {\kern 1pt} {\kern 1pt} {\kern 1pt} {\kern 1pt} {\kern 1pt} aP{K_{c1}} - aP{K_{c1}}^*\\
 = {\kern 1pt} {\kern 1pt} {\kern 1pt} (V - T - \gamma {P_{pub}}) - (V - {T^*} - {\gamma ^*}{P_{pub}})\\
 = {\kern 1pt} {\kern 1pt} ({T^*} - T) + ({\gamma ^*} - \gamma ){P_{pub}}
\end{array}\]

\emph{\textbf{Probability Analysis:}} For above discussions, there are three situations leading to the aborting of ${\cal C}$:
\begin{itemize}
  \item $E_1$: ${\cal F}$ asks the private key queries of the challenged identity $I{D_\alpha }$.
  \item $E_2$: ${\cal F}$ does not choose $I{D_\alpha }$ as the sender's identity in the challenge phase.
  \item $E_3$: ${\cal C}$ aborts in a signcryption query due to the collision on hash operation.
\end{itemize}

Only if ${\cal C}$ does not reject the game, the DLP can be solved. As a result, the probability of ${\cal C}$ not aborting is: $\Pr [\neg abort] = \Pr [\neg {E_1} \wedge \neg {E_2} \wedge \neg {E_3}]$.

\begin{figure}[htpt]
   %Requires \usepackage{graphicx}
\centering
  \includegraphics[ width=9cm ]{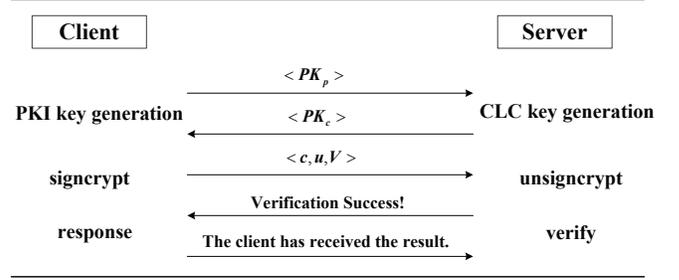}
  \caption{The structure diagram of the interprocess communication.}
\end{figure}

We know that $\Pr [{E_1}] = \frac{{{q_{ep}}}}{{{q_1}}}$, $\Pr [{E_2}] = (1-\frac{1}{{{q_1}}})$, and $\Pr [{E_3}] \le \frac{{{q_s}}}{{{2^l}}}$. Therefore, $\Pr [abort] \ge \frac{1}{{{q_1}}}(1 - \frac{{{q_{ep}}}}{{{q_1}}})(1 - \frac{{{q_s}}}{{{2^l}}})$.

In addition, the probability that ${\cal F}$ correctly guesses the hash value of ${H_1}$ is $\frac{1}{{{q_1}}}$. In conclusion, we have
${\xi ^*} \ge \frac{\xi }{{{q_1}^2}}(1 - \frac{{{q_{ep}}}}{{{q_1}}})(1 - \frac{{{q_s}}}{{{2^l}}})$.

The security proof of CPHS is similar to that of PCHS.

\section{Performance Analysis}
In this section, we assume that the client is in 5G Mobile Internet slicing (PKI environment) and the server is in 5G Vehicle Internet slicing (CLC environment). To simulate the interprocess communication between the client and the server in different public cryptosystems. We set up two Raspberry Pis as the experimental platform. The structure diagram of the interprocess communication is shown in Fig. 4. and the simulation results are shown in Fig. 5.

\begin{table*}
\scriptsize \caption{Performance comparison of each signature scheme}\label{security
features}
\begin{center}
\begin{tabular}{c c c c c c c c}
 %The scheme&$in \cite{He}$&$He[2]$&$in %\cite{Goriparthi}$&$in \cite{Tsu-Yang}$&$scheme$ \\
\toprule
    The scheme&Key

Generation&Signcryption&Unsigncryption&Communication cost&Communication Direction \\
\midrule
  LZT-I[24]  &4S+H&P+3S+3H+E&3P+2H+E& 2$|{G_1}|$+$|m|$ &PKI $\rightarrow$ IBC   \\

  LZT-II[24] &4S+H&P+2S+2H+E&3P+S+3H+E&2$|{G_1}|$+$|m|$ &IBC $\rightarrow$ PKI  \\

  LHJ[25] &4S+2H&P+3S+3H+E&5P+3H&2$|{G_1}|$+$|m|$ &CLC $\rightarrow$ IBC  \\

  ZZW[26] &P+5S+2H&P+3S+2H+2E&2P+2S+3H+E&3$|{G_1}|$+$|ID|$+$|m|$ &IBC $\rightarrow$ CLC\\

  WLZ[27] &2P+7E&3S+H+4E&5P+2S+H+E&3$|{G_1}|$+$|{G_2}|$ &IBC $\rightarrow$ PKI\\

  PCHS       &4S+H&4S+2H&3S+2H&$|{G_1}|$+$|{Z_q^*}|$+$|m|$ &PKI $\rightarrow$ CLC\\

  CPHS       &4S+H&3S+2H&4S+2H&$|{G_1}|$+$|{Z_q^*}|$+$|m|$ &CLC $\rightarrow$ PKI\\
\bottomrule
\end{tabular}
\end{center}
\end{table*}\

\begin{figure}[htpt]
\centerline{
\subfigure[The client]{\includegraphics[width=1.7in,height=1.3in]{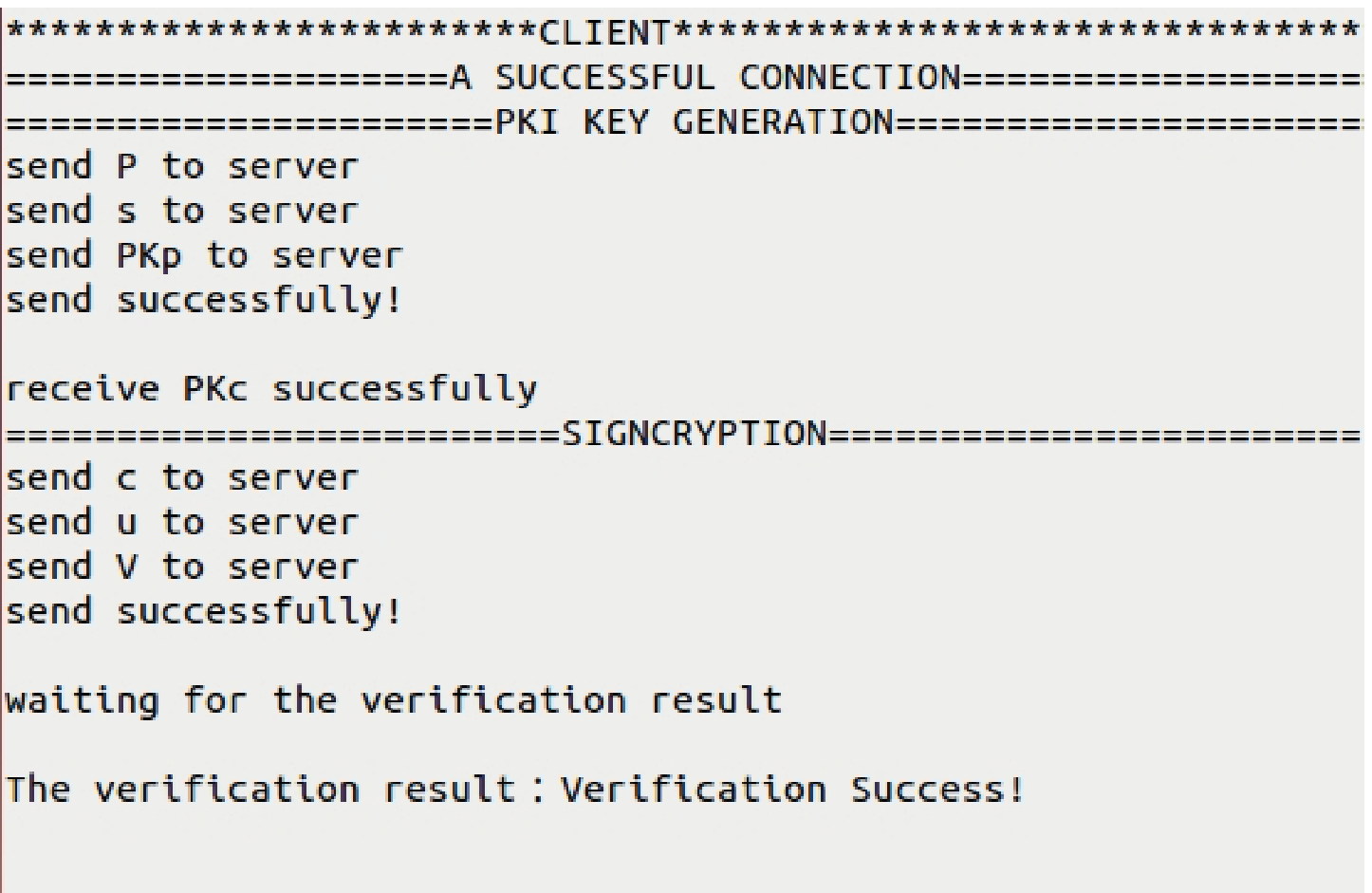}} %
%\hfil
%\hspace{-3mm}
%\vspace{1cm}
\subfigure[The server]{\includegraphics[width=1.7in,height=1.3in]{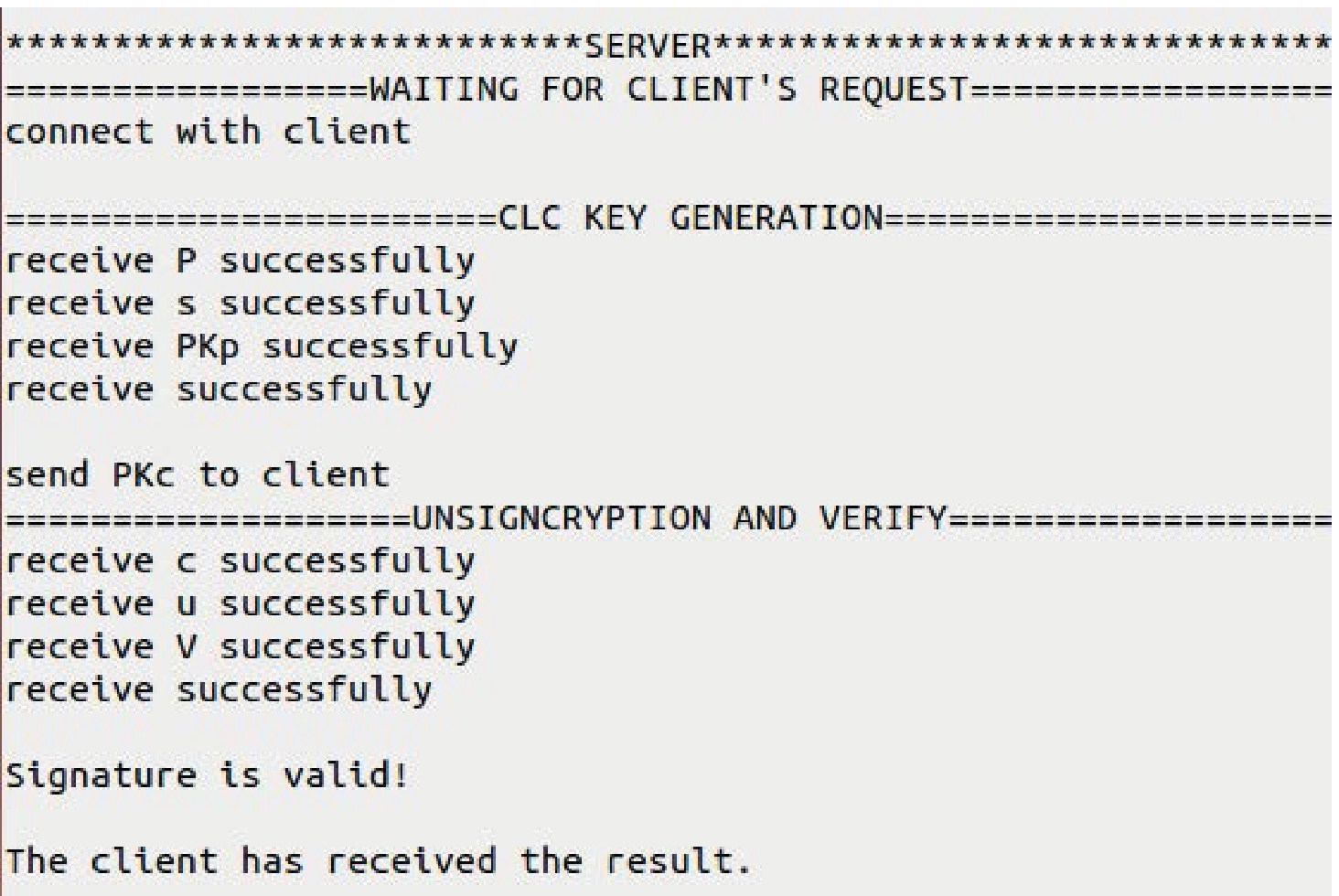}} %
%\hfil
%\hspace{-3mm}
}
%%%\centerline
% \captionsetup{skip=-1pt}
%%%\setlength{\abovecaptionskip}{-0.5 mm}
\setlength{\belowcaptionskip}{-2 mm}
\caption{The simulation result between the client and the server.}
\label{Simulation_Time}
\vspace{-1mm}
\end{figure}

\begin{figure*}[htpt]
\centerline{
  \subfigure[Time comparison on key generation]{\includegraphics[width=2.3in]{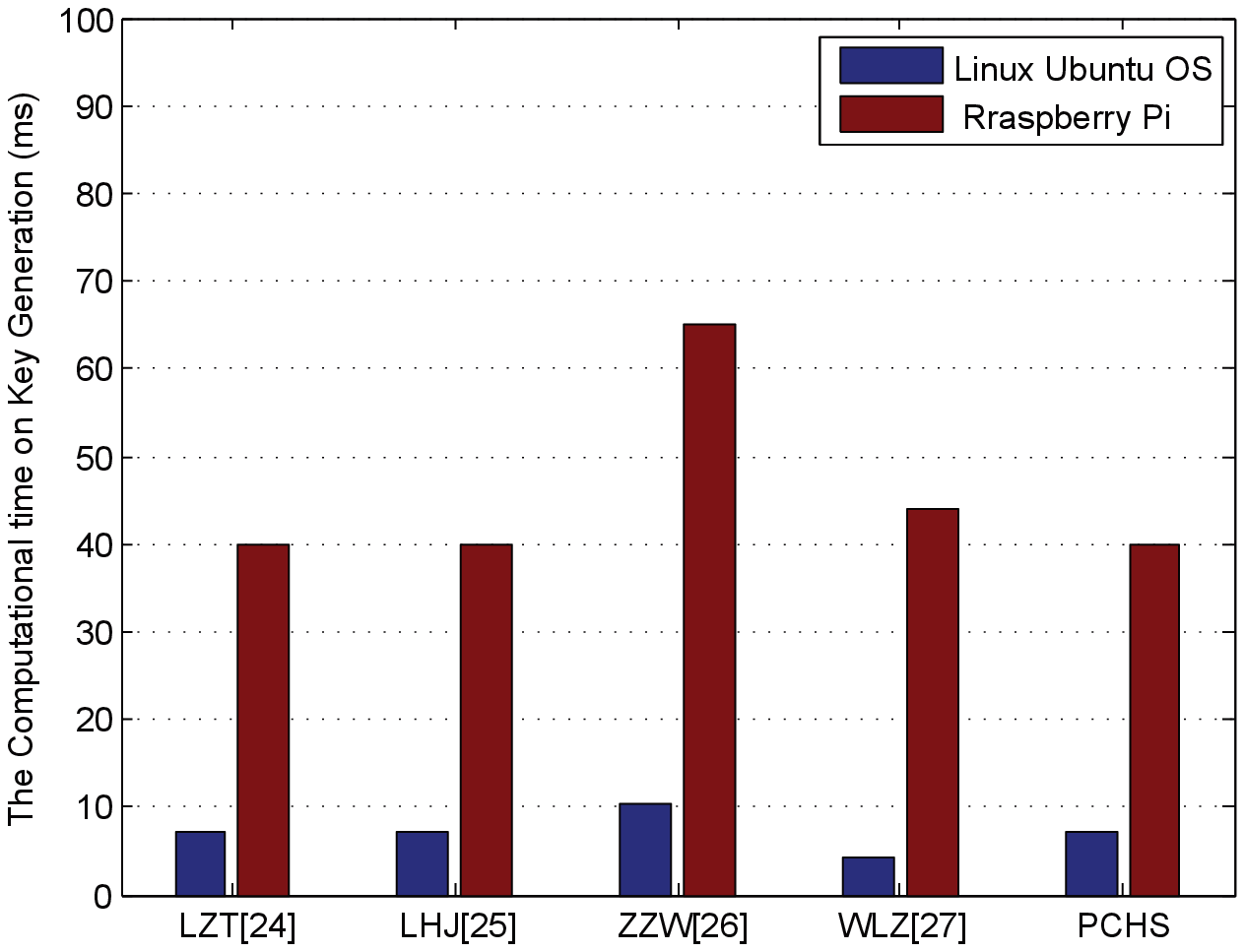}}
  \subfigure[Time comparison on signcryption and unsigncryption]{\includegraphics[width=2.3in]{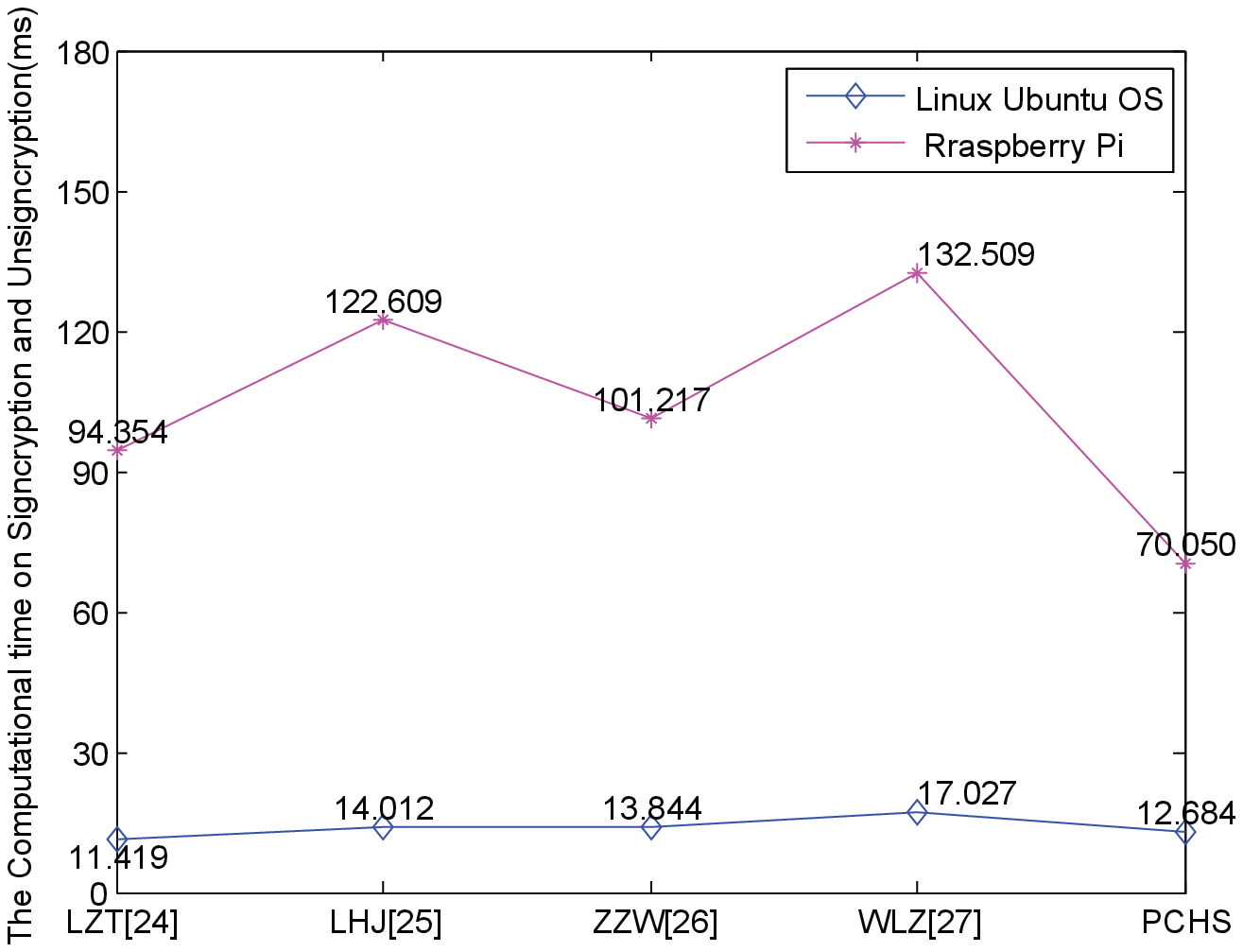}}
  \subfigure[Time comparison in total]{\includegraphics[width=2.3in]{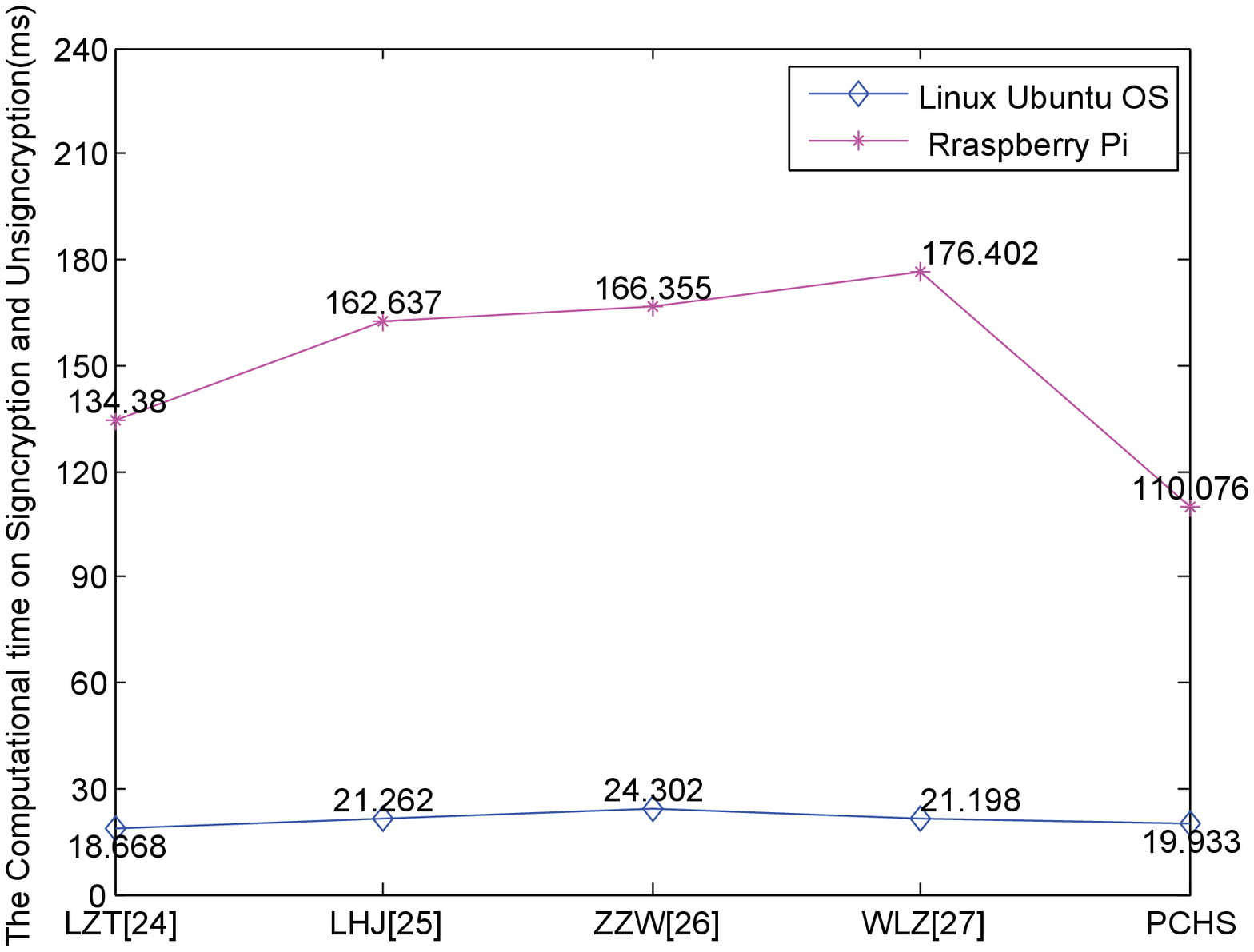}}}
\setlength{\belowcaptionskip}{-2 mm}
\caption{Comparison of time consumption among different schemes.}
\label{Simulation_Time1}
\vspace{-1mm}
\end{figure*}
In the first step, the client generates the public key ${PK_p}$ and sends it to the server. Also the server receives his/her public key ${PK_c}$ to the client. In the second step, via the signcryption, the client generates the ciphertext $\sigma $ and sends it to the server. Then, the server unsigncrypts and verifies the ciphertext. If it is verified, the server sends ``Verification Success!'' to the client. In the last step, if the client obtains the response from the server, it sends ``The client has received the result.'' to the server as a reply.

For the theoretical complexity analysis, we compare our schemes with several existing schemes [24], [25], [26] and [27] in Table I. Let $\left| {{G_1}} \right|$ and $\left| {{G_2}} \right|$ denote the length of the elements in group $G_1$ and $G_2$ respectively, $\left| Z_q^* \right|$ denote the length of an element in field $Z_q^*$, $\left| m \right|$ denote the length of a message, $\left| {ID} \right|$ denote the the length of a user's identity. $S$ is the scalar multiplication in ${G_1}$, $E$ means the exponentiation in ${G_2}$, ${P}$ denotes the bilinear pairing, ${H}$ is the hash operation in $Z_q^*$.

From Table I, we can clearly find that all schemes except ours use bilinear pairings and exponentiation computations in the signcryption and unsigncryption stages. Furthermore, compared to the schemes in [24, 25, 26], the scheme [27] involves more bilinear pairings, which lead to more time consumption. As for the communication overhead, WLZ [27] is the highest among these schemes because one of its ciphertext elements is in group $G_2$. Observing the last column in Table I, we find that the schemes in [25, 26, 27] are one-way communications, which have the limitations in practice. The schemes in [24-27] are only suit for the heterogeneous environments of PKI-to-IBC or IBC-to-CLC. Our schemes are designed specially for the mutual heterogeneous communication of PKI-to-CLC public cryptosystems innovatively.

For the quantitative analysis, we set up two different environments for simulation. The first experimental platform is in Ubuntu OS over an Inter Pentium 2.70 GHz dual core processor and 1024 MB memory. The second one is in Raspberry Pi 3B+. The cryptography library we used is PBC-0.5.14. We choose the type A curve ${y^2} = {x^3} + x$ in simulation because the math algorithm is more efficient on exponentiation in $G_2$. We first assess the cryptographic operations, and each operation is run 10000 times to eliminate the influence of random disturbance. The detailed results are shown in Table II. Generally, the time overhead in Raspberry Pi is about ten times higher than that in Ubuntu. Furthermore, we also find that the time cost on the scalar multiplication is higher than that on the bilinear pairing in Ubuntu environment, while the result is opposite in Raspberry Pi platform, which would lead to the significant changes among schemes in the two simulation environments.
\begin{table}
\centering
\caption{Operating Time in different simulation environments(ms)}
\begin{tabular}{|c|c|c|c|c|} \hline
\backslashbox{Environment}{
Operation} & P & S & E & H \\
\hline
Ubuntu OS &1.396&1.812&0.197&0.001\\ \hline
Raspberry Pi &15.104&10.006&1.955&0.002\\
\hline
\end{tabular} \\
\end{table}

Before running the schemes, we first test the time consumption on their private/public key pair generation. As shown in Fig. 6(a), although the time cost on key generation of each scheme is very close, WLZ [27] achieves the lowest computation cost in this part because there is no scalar multiplication but only a few bilinear pairings. Fig. 6(b) shows the time consumption on the signcryption and unsigncryption among different schemes in two platforms. Fig. 6(c) indicates the total overhead of the key generation, the signcryption and the unsigncryption. As a whole, ZZW [26] has the highest computational overhead in Ubuntu environment, while WLZ [27] takes the highest cost in Raspberry Pi platform. Without any bilinear pairing, our scheme is more efficient on the computational overhead.

\section{Conclusion}
In this paper, we propose two signcryption schemes without the bilinear pairing between PKI and CLC cryptosystems, which can achieve secure mutual heterogeneous communications of 5G network slicings. When the users in the 5G slicing based on PKI environment (such as a Mobile Internet slicing) try to communicate with the users in the 5G slicing based on CLC (such as a Vehicle Internet slicing), our schemes can provide the available and efficient solutions. Meanwhile, they have the IND-CCA2 under the CDHP and the EUF-CMA under the DLP in the random oracle model. We also simulate the interprocess communication between two different public cryptosystems on Raspberry Pi. Compared to the existing schemes, our schemes are innovative and more efficient for heterogeneous communications of 5G network slicings.

\end{document}